\documentclass[a4paper,notitlepage,eqsecnum,nofootinbib]{revtex4-1}
\usepackage{slashed,bbold,bm,amsmath,amssymb,mathtools,wrapfig,epsfig} 
\usepackage[usenames]{color}
\usepackage{hyperref}
\usepackage{tcolorbox}

\hypersetup{
    colorlinks=true,       
    linkcolor=red,          
    citecolor=blue,        
    filecolor=green,      
    urlcolor=magenta           
}

\newcommand{\crt}{\\[2mm]}
\newcommand{\nn}{\nonumber}

\newcommand{\mrm}[1] {{\mathrm{#1}}}

\newcommand{\bs}[1]{\boldsymbol{#1}}

\newcommand{\eps}{\epsilon}
\newcommand{\veps}{\varepsilon}

\newcommand{\vphi}{\varphi}
\newcommand{\order}[1]{${\cal O}\left(#1 \right)$}

\newcommand{\eq}[1]{(\ref{#1})}
\newcommand{\fig}[1]{Fig.~\ref{#1}}

\newcommand{\halft}{{\textstyle \frac{1}{2}}}

\newcommand{\sfrac}[2]{{\textstyle\frac{#1}{#2}}}

\newcommand{\ket}[1]{\left\vert{#1}\right\rangle}

\newcommand{\com}[2]{\left[{#1},{#2}\right]}

\newcommand{\acom}[2]{\left\{{#1},{#2}\right\}}

\newcommand{\xv}{{\bs{x}}}

\newcommand{\gz}{\gamma^0}
\newcommand{\go}{\gamma^1}

\newcommand{\qb}{{\bar{q}}}

\newcommand{\kum}{{\,{_1}F_1}}

\newcommand{\phip}{\phi^{\scriptscriptstyle{(P)}}}

\newcommand{\phii}{\phi^{\scriptscriptstyle{(\infty)}}}

\newcommand{\Phip}{\Phi^{\scriptscriptstyle{(P)}}}

\newcommand{\dx}{\partial_x}

\newcommand{\dm}{{\Delta m^2}}

\begin{document}

\title{Dirac states from the 't Hooft model}

\author{Paul Hoyer}
\affiliation{ \vspace{1mm} Department of Physics, POB 64, FIN-00014 University of Helsinki, Finland}
\email{paul.hoyer@helsinki.fi} 

\begin{abstract}  

The dynamics of a light fermion bound to a heavy one is expected to be described by the Dirac equation with an external potential. The potential breaks translation invariance, whereas the bound state momentum is well defined. Boosting the bound state determines the frame dependence of the light fermion dynamics. I study the Dirac limit of
QCD$_2$ in the limit of $N_c \to \infty$.
The light quark wave function turns out to be independent of the frame of the bound state, up to an irrelevant Lorentz contraction. The discrete bound state spectrum determines corresponding discrete energies of the Dirac equation, which for a linear potential allows a continuous spectrum.

\end{abstract}

\maketitle

\vspace{-.5cm}

\parindent 0cm

\section{Introduction} \label{secI}

The Dirac equation \cite{Dirac:1928hu,*Dirac:1928ej} for an electron of mass $m_1$ in a Coulomb potential can be derived by summing Feynman ``ladder diagrams'', where photons are exchanged between the electron and a heavy particle \cite{Brodsky:1971we,Gross:1982nz,Neghabian:1983vm}. As the mass $m_2$ of the heavy particle tends to infinity in its rest frame the electron experiences a static Coulomb potential. Exchanges with crossed photon lines ($Z$-diagrams) generate the negative energy components of the Dirac wave function.

The Dirac equation with a static potential $V(\xv)$ gives eigenstates of energy but not of spatial momentum, because the potential breaks space translation invariance. On the other hand, the bound state of a light and a heavy particle can have a well-defined center-of-mass (CM) momentum. Provided $m_2 \gg V(\xv)$ the light particle dynamics in the bound state should be governed by the Dirac equation. Bound states with non-vanishing CM momentum allows to view the Dirac dynamics in any reference frame.

Boosts of canonically quantized (equal-time) bound states involve interactions, unlike the standard covariance of the Dirac equation. Boost generators do not commute with the Hamiltonian in the Poincar\'e Lie algebra, implying that the dynamics is frame dependent. The example of the Hydrogen atom, described by the Schr\"odinger equation in the rest frame, shows that bound state boosts are non-trivial \cite{Jarvinen:2004pi}. Relativistically bound, Poincar\'e covariant bound states can generally not be studied analytically.

Here I consider QCD in $D=1+1$ dimensions (QCD$_2$) in the limit of a large number of colors, $N_c \to \infty$. As first noted by 't Hooft \cite{tHooft:1974pnl}, quark pair production is suppressed in this limit, allowing an exact solution for the $q\qb$ Fock states. 't~Hooft used light-front (LF) quantization at equal $x^+=t+z$. The LF method neglects ``zero-modes'', by assuming particles to carry positive LF momenta ($p^+>0$). Consequently, 't Hooft required that the LF wave functions vanish at quark momentum fraction $x_B=0$. Bars and Green \cite{Bars:1977ud} were able to reproduce the results of 't Hooft using quantization at equal time $t$. They needed to modify the Feynman $i\eps$ prescription of the quark propagators, but Poincar\'e symmetry was restored in the $q\qb$ sector. Numerous later studies were based on their methods.

A recent study of QCD$_2$ at $N_c \to \infty$ \cite{Hoyer:2025cpf} started from canonically quantized $q\qb$ states, avoiding the use of singular quark propagators. This method agreed with 't Hooft's equation in the infinite momentum frame. However, the wave function did not vanish at $x_B=0$, and included contributions with negative kinetic energy. This implied that the bound state spectrum differed from that of 't Hooft.

As argued above, the dynamics of a light constituent in a bound state should be described by the Dirac equation with a Coulomb potential, in the limit where the other constituent is heavy. A previous study of this limit \cite{Kalashnikova:1997tb}, using a method based on that of Bars and Green \cite{Bars:1977ud}, gave non-local interactions with scalar and space-like vector potentials. Here I find that the bound states defined as in \cite{Hoyer:2025cpf} reduce to the usual Dirac equation with a linear Coulomb potential.

The Dirac equation with a linear potential has some unusual features \cite{Plesset:1930zz}. The wave function oscillates with frequency $\propto x$ and constant norm as $x \to \infty$. Similarly as for plane waves the solutions allow a continuum of energy eigenvalues. At large $x$ the potential $V'|x|$ is balanced by a correspondingly large and negative kinetic energy of the fermion. This describes a positron being repelled by the potential \cite{Hoyer:2021adf}. Viewing the Dirac states as the $m_2 \to \infty$ limit of $q\bar q$ bound states adds insight into their spectrum and frame dependence.

\section{The Dirac equation from $\mathbf{q\qb}$ bound states} \label{secII}

\subsection{Rest frame} \label{secII.A}

In temporal $(A_a^0=0)$ gauge the rest frame $q\qb$ states of QCD$_2$
can at time $t=0$ be expressed as \cite{Hoyer:2025cpf}
\begin{align}
\ket{M}&= \int dx_1 dx_2\,\bar\psi_1(x_1)\mathbb{P}(x_2 \to x_1)\,\Phi(x_1-x_2)\psi_2(x_2)\ket{0} \hspace{1cm} (t=0)  \label{2.1} \crt
&\mathbb{P}(x_2 \to x_1) \equiv \mathbb{P}\exp\Big[ig\int_{x_2}^{x_1}dx\,A^1_a(x)T^a\Big]
 \hspace{2cm} (\mathbb{P}:\ \mbox{path ordering}) \label{2.2}
\end{align}
Here $M$ is the bound state mass, $\psi_1$ is the field of the quark with mass $m_1$ (having two Dirac and $N_c$ color components), and $\psi_2$ is the field of the quark with mass $m_2$. The gauge link $\mathbb{P}(x_2 \to x_1)$ ensures that $\ket{M}$ is invariant under time independent gauge transformations (which preserve $A^0_a=0$).  For $\ket{M}$ to be an eigenstate of the Hamiltonian the $c$-numbered, color singlet wave function $\Phi(x)$ should satisfy the Bound State Equation (BSE),
\begin{align} \label{2.3}
i\dx\acom{\alpha_1}{\Phi(x)}+m_1\gz\Phi(x)-m_2\Phi(x)\gz +V(x)\Phi(x) = M\Phi(x)\,; \hspace{1.cm} V(x) = V'|x|
\end{align}  
where $\alpha_1=\gz\go$ and $V'=\halft g^2C_F$. In $D=1+1$ dimensions we may represent the Dirac matrices by Pauli matrices,
\begin{align} \label{2.4}
\gz = \sigma_3,\ \ \go = i\sigma_2,\ \ \alpha_1 = \sigma_1
\end{align}
In the $N_c \to \infty$ limit $g \propto 1/\sqrt{N_c} \to 0$  \cite{tHooft:1974pnl}. Hence $q\qb$ pair production is suppressed, ensuring that the BSE is exact at fixed $V'$. The analytic solutions of \eq{2.3} were found in \cite{Dietrich:2012un} \footnote{The energy scale was fixed by $V'=\halft$ in \cite{Dietrich:2012un}. Here I shall keep the dependence on $V'$, which means that powers of $2V'$ need to be inserted in the dimensionful quantities of \cite{Dietrich:2012un}.}. 

The four Dirac components of $\Phi(x)$ may be expressed in terms of the scalar functions $\phi_j(x),\ (j=0,1,2,3)$,
\begin{align} \label{2.5}
\Phi(x) = \phi_0(x) +\sigma_1\,\phi_1(x) +i\sigma_2\phi_2(x)+\sigma_3\phi_3(x)
\end{align}
The BSE \eq{2.3} then implies that $\Phi$ may be expressed in terms of just $\phi_0$ and $\phi_1$,
\begin{align} \label{2.6}
\Phi(x) &= \Big[1+\frac{m_1-m_2}{M-V}\,\sigma_3\Big]\phi_0(x) + \Big[\sigma_1+\frac{m_1+m_2}{M-V}\,i\sigma_2\Big]\phi_1(x) \nn\crt
2i\dx\phi_1 =(M-V)&\Big[1- \frac{(m_1-m_2)^2}{(M-V)^2}\Big]\phi_0 \hspace{2cm}
2i\dx\phi_0 =(M-V)\Big[1- \frac{(m_1+m_2)^2}{(M-V)^2}\Big]\phi_1
\end{align}
In the limit of large $m_2$ I define the Dirac bound state energy $M_D$ by
\begin{align} \label{2.7}
M \buildrel {m_2\to\infty}\over{\rule{0pt}{2mm}=} m_2+M_D
\end{align} 
For $m_2 \to \infty$ at fixed $m_1,\,M_D$ and $x$ \eq{2.6} becomes
\begin{align} \label{2.8}
\lim_{m_2\to\infty}\Phi & = (1-\sigma_3)\phi_0 + (\sigma_1+i\sigma_2)\phi_1
=2\left(\begin{array}{cc}0 & \phi_1 \crt 0 & \phi_0 \end{array} \right) 
= 2\sigma_3\left(\begin{array}{c} \phi_1 \crt -\phi_0 \end{array} \right)
\left(\begin{array}{cc} 0 & 1\end{array} \right) \nn\crt
i\dx\phi_1 &\buildrel {m_2\to\infty}\over{\rule{0pt}{2mm}=} (M_D-V+m_1)\phi_0 \hspace{2cm}
i\dx\phi_0 \buildrel {m_2\to\infty}\over{\rule{0pt}{2mm}=} (M_D-V-m_1)\phi_1
\end{align}
Hence at large $m_2$ the spinor with components $\phi_1$ and $-\phi_0$ satisfies the Dirac equation, with fermion mass $m_1$, potential $V=V'|x|$ and energy $M_D$,
\begin{align} \label{2.9}
(-i\sigma_1\dx + m_1\sigma_3+V)\left(\begin{array}{c} \phi_1 \crt  -\phi_0 \end{array} \right)
= M_D\left(\begin{array}{c} \phi_1 \crt  -\phi_0 \end{array} \right)
\end{align}
Non-leading contributions of \order{V/m_2} to \eq{2.8} imply that $V \ll m_2$ is required for the Dirac limit.

\subsection{CM momentum $P>0$} \label{secII.B}

Eigenstates of momentum $P^1 \equiv P>0$ at $t=0$ may in QCD$_2$
 be expressed similarly to \eq{2.1} \cite{Hoyer:2025cpf},
\begin{align} \label{2.10}
\ket{M,P}&= \int dx_1 dx_2\,\bar\psi(x_1)\mathbb{P}(x_2 \to x_1)\,\exp\big[\halft iP(x_1+x_2)+i\vphi(x_1-x_2)\big]\Phip(x_1-x_2)\psi(x_2)\ket{0} 
\end{align}
As noted in \cite{Dietrich:2012un} it is convenient to separate a phase $\vphi(x)$ from the wave function $\Phip(x)$, with
\begin{align} \label{2.11}
\vphi(x) \equiv \veps(x)\,\frac{m_1^2-m_2^2}{4V'}\,\log\left[\left(\frac{M}{E+P}\right)^2 \left|\frac{E-V+P}{E-V-P}\right|\right]
\end{align}
where $\veps(x\,\genfrac{}{}{0pt}{2}{>}{<}\, 0)=\pm 1$ and $E=\sqrt{M^2+P^2}$\,. Note that $\vphi(0)=0$ for all $P$, and $\vphi(x)=0$ for $P=0$. For $m_2 \to \infty$ with $M=m_2+M_D$ as in \eq{2.7} the phase factor in \eq{2.10} becomes
\begin{align} \label{2.12}
\lim_{m_2\to\infty}\exp\big[\halft iP(x_1+x_2)+i\vphi(x_1-x_2)\big]=\exp(iP\,x_2)
\end{align}
showing that the CM momentum $P$ of the bound state is carried by the heavy quark.
The state \eq{2.10} is an eigenstate of the QCD$_2$ Hamiltonian with eigenvalue $E$ if $\Phip(x)$ satisfies the BSE,
\begin{align} \label{2.13}
i\dx\acom{\sigma_1}{\Phip(x)}-(\dx\vphi)\acom{\sigma_1}{\Phip(x)}-\halft P\com{\sigma_1}{\Phip(x)}+m_1\sigma_3\Phip(x)-m_2\Phip(x)\sigma_3 = (E-V)\Phip(x)
\end{align}  
where
\begin{align} \label{2.14}
\dx\vphi(x) = \halft P\,\frac{m_1^2-m_2^2}{M^2-2EV+V^2}
\end{align}
Using the expansion \eq{2.5} for $\Phip(x)$ in terms of the scalar functions $\phip_j(x)$ the BSE implies
\begin{align}
\Phip(x) &= \Big[1+(m_1-m_2)\frac{(E-V)\sigma_3 + P\,i\sigma_2}{(E-V)^2-P^2}\Big]\phip_0(x) +
\Big[\sigma_1+(m_1+m_2)\frac{P\sigma_3 + (E-V)\,i\sigma_2}{(E-V)^2-P^2}\Big]\phip_1(x)  \label{2.15}\crt
2i\dx\phip_1 &=(E-V)\Big[1- \frac{(m_1-m_2)^2}{(E-V)^2-P^2}\Big]\phip_0 \hspace{2cm}
2i\dx\phip_0 =(E-V)\Big[1- \frac{(m_1+m_2)^2}{(E-V)^2-P^2}\Big]\phip_1  \label{2.16}
\end{align}
Lorentz invariance requires $E=\sqrt{M^2+P^2}$\,. The $P$-dependence of the $\phip_{j}(x)$ is then determined by \eq{2.16}. 
In a boost from the rest frame with Lorentz factor $\gamma$,
\begin{align} \label{2.17}
E &= \gamma M \hspace{2cm} P = \sqrt{\gamma^2-1}\,M \hspace{2cm} M=m_2+M_D
\end{align}
the Lorentz contraction of $x$ can be taken into account through the variable $u$,
\begin{align} \label{2.18}
u &= \gamma\,x \hspace{2cm} \dx = \gamma\,\partial_u
\end{align}
For $m_2 \to \infty$ at fixed $M_D$, $\gamma$ and $u$ the BSE \eq{2.16} reduces to the Dirac equation \eq{2.8} in the variable $u$,
\begin{align} \label{2.19}
i\partial_u\phip_1 & = (M_D-V'|u|+m_1)\phip_0 \hspace{2cm}
i\partial_u\phip_0 = (M_D-V'|u|-m_1)\phip_1
\end{align}
Since Lorentz contraction is irrelevant for a single particle the Dirac equation is in fact frame independent.

\section{Solutions of the Bound State and Dirac equations} \label{secIII}

\subsection{The Bound State Equation} \label{secIIIA}

The form of the BSE \eq{2.16} suggests to describe the $x$-dependence at fixed $P$ using the variable $\tau_P(x)$,
\begin{align} \label{3.1}
\tau_P(x) &\equiv \big[(E-V)^2-P^2\big]/V' \hspace{3cm}
\dx = \frac{\partial\tau_P}{\partial x}\partial_{\tau_P} = -2\veps(x)(E-V)\,\partial_{\tau_P}
\end{align}
Expressed in terms of $\tau_P$ \eq{2.16} does not explicitly depend on $P$ nor $E$ (at any value of $m_1,m_2$ and $x \geq 0$),
\begin{align} \label{3.2}
-4i\partial_{\tau_P}\phip_1 &=\Big[1- \frac{(m_1-m_2)^2}{V'\tau_P}\Big]\phip_0 \hspace{2cm}
-4i\partial_{\tau_P}\phip_0 =\Big[1- \frac{(m_1+m_2)^2}{V'\tau_P}\Big]\phip_1 
\end{align}
Given the solutions $\phip_1(\tau_P)$ and $\phip_0(\tau_P)$ their $x$-dependence at any $P$ is determined by $\tau_P(x)$ \eq{3.1}. The $2\times 2$ wave function $\Phip$ \eq{2.15} depends also explicitly on $E$ and $P$. Since $V(x)=V'|x|$ is non-analytic at $x=0$ we may first solve \eq{2.16} for $x\geq 0$ and then determine $\phip_{0,1}(x<0)$ using the $x \to -x$ symmetry of the BSE. This imposes a smoothness condition on $\phip_{0,1}(\tau_P)$ at $x=0$, which requires that $\tau_P(x=0) = (E^2-P^2)/V'$ is independent of $P$. There are regular solutions at all $x$ and $P$ only if $E^2-P^2=M^2$, as required by Lorentz covariance.

Denoting $\dm \equiv m_1^2-m_2^2$ the general solution of \eq{3.2} given in \cite{Dietrich:2012un} is
\begin{align} \label{3.3}
\phip_1(\tau_P)+\phip_0(\tau_P) = e^{i\tau_P/4}&\big[(a+ib)m_1|\tau_P|^{-i\dm/4V'}\kum(im_2^2/2V',1-i\dm/2V',-i\tau_P/2) \nn\crt
&\hspace{-2mm}-(a-ib)m_2|\tau_P|^{i\dm/4V'}\kum(im_1^2/2V',1+i\dm/2V',-i\tau_P/2)\big]\nn\crt
\phip_1(\tau_P)-\phip_0(\tau_P) = e^{-i\tau_P/4}&\big[(a-ib)m_1|\tau_P|^{i\dm/4V'}\kum(-im_2^2/2V',1+i\dm/2V',i\tau_P/2) \nn\crt
&\hspace{-3mm}-(a+ib)m_2|\tau_P|^{-i\dm/4V'}\kum(-im_1^2/2V',1-i\dm/2V',i\tau_P/2)\big]
\end{align}
where $a$ and $b$ are complex constants. The $2\times 2$ wave function \eq{2.15} can be singular at $\tau_P=0$. Setting $a+ib=0$ in \eq{3.3} avoids a $1/\tau_P$ pole in the $P \to \infty$ limit \cite{Dietrich:2012un}. Using $\kum(b-a,b,z) = e^z\kum(a,b,-z)$ then gives
\begin{align} \label{3.4}
\phip_1(\tau_P)&= \halft(a-ib)e^{-i\tau_P/4}\,|\tau_P|^{i\dm/4V'}\nn\crt
&\times\big[m_1\kum(-im_2^2/2V',1+i\dm/2V',i\tau_P/2)-m_2\kum(1-im_2^2/2V',1+i\dm/2V',i\tau_P/2)\big] \nn\crt
\phip_0(\tau_P)&= -\halft(a-ib)e^{-i\tau_P/4}\,|\tau_P|^{i\dm/4V'}\nn\crt
&\times\big[m_1\kum(-im_2^2/2V',1+i\dm/2V',i\tau_P/2)+m_2\kum(1-im_2^2/2V',1+i\dm/2V',i\tau_P/2)\big]
\end{align}

If $\Phip(x)$ is a solution of the BSE \eq{2.13} at any $P$ then so is ${\Phip}^*(-x)$. With \eq{3.4} for $x\geq 0$ we may define 
\begin{align} \label{3.6}
\Phip(-x) &= \eta {\Phip}^*(x) \hspace{1cm} (\eta=\pm 1)
\end{align}
This means that $\Phip(x=0)$ is real ($\eta=1$) or imaginary ($\eta=-1$). In either case the phase difference between $\phip_0$ and $\phip_1$ can only be $0$ or $\pi$ at $x=0$, so
\begin{align} \label{3.7}
\mrm{Im}&\big[\phip_0(x=0)\, {\phip_1}^*(x=0)\big] =0
\end{align}
Up to an overall normalization, the complex parameter $a-ib$ in \eq{3.3} may be fixed by setting $\phip_1(x=0)=1 \ (i)$ for $\eta=1\ (-1)$. The phase condition \eq{3.7} on $\phip_0(x=0)$ then determines the discrete bound state masses $M$.

For $P \to \infty$ at fixed $u$ (defined as in \eq{2.18}) also $\tau_P \simeq M^2/V' -2M u$ is fixed. In this limit the wave function becomes, using $\kum(a,b,z)-\kum(a-1,b,z)=(z/b)\kum(a,b+1,z)$,
\begin{align} \label{3.5}
\lim_{P\to\infty}&\Phip(u) = \frac{\gamma}{M-2V'|u|}(\sigma_3+i\sigma_2)\big[m_1(\phii_1+\phii_0)+m_2(\phii_1-\phii_0)\big] \nn\crt
&= -i(a-ib)\frac{\gamma\, m_1m_2 M}{2V'+i\dm}\,(\sigma_3+i\sigma_2)e^{-i\tau_P/4}\,|\tau_P|^{i\dm/4V'}
\kum(1-im_2^2/2V',2+i\dm/2V',i\tau_P/2)
\end{align}

\subsection{The Dirac equation} \label{secIIIB}

The Dirac equation in $D=1+1$ dimensions with a linear potential $V=V'|x|$ is
\begin{align} \label{3.8}
(-i\sigma_1\dx + m\sigma_3+V'|x|)\Psi_D= M_D&\Psi_D \hspace{2cm}
\Psi_D = \left(\begin{array}{c} \vphi_D \crt  \chi_D \end{array} \right) \nn\crt
-i\dx\vphi_D = (M_D-V+m)\chi_D &\hspace{2cm} -i\dx\chi_D = (M_D-V-m)\vphi_D \nn\crt
\dx^2\vphi_D + \frac{V'\veps(x)}{M_D-V+m}\,\dx&\vphi_D + \big[(M_D-V)^2-m^2\big]\vphi_D =0 
\end{align}
For $x \to \infty$ the last equation reduces at leading order to $(\dx^2+V^2)\vphi_D=0$, so $\vphi_D(x\to\infty) \sim \exp(\pm i V'x^2/2)$. This allows a continuum of eigenvalues $M_D$ \cite{Plesset:1930zz}, like for plane waves. The bound state spectrum discussed above in Section \ref{secIIIA} is, on the other hand, discrete. Hence obtaining the Dirac equation as a limit of a bound state with a heavy constituent determines a discrete spectrum also for the Dirac equation. 

An analytic solution of the Dirac equation \eq{3.8} for $x\geq 0$ was given in \cite{Dietrich:2012un}. For any complex constants  $c_1,c_2$,
\begin{align} \label{3.9}
\vphi_D(x)+\chi_D(x)&= e^{-i(M_D-V)^2/2V'}\big\{c_1\kum[-im^2/4V',\halft,i(M_D-V)^2/V'] \nn\crt
&\hspace{2.6cm}+c_2m(M_D-V)\kum[\halft-im^2/4V',\sfrac{3}{2},i(M_D-V)^2/V']\big\} \nn\crt
\vphi_D(x)-\chi_D(x)&= -i\,e^{-i(M_D-V)^2/2V'}\big\{c_1 m(M_D-V)\kum[1-im^2/4V',\sfrac{3}{2},i(M_D-V)^2/V'] \nn\crt
&\hspace{3.1cm}+c_2\kum[\halft-im^2/4V',\halft,i(M_D-V)^2/V']\big\}
\end{align}
If $\Psi_D(x)$ is a solution of the Dirac equation \eq{3.8}, then so is $\Psi_D^*(-x)$. As in \eq{3.6} the solution for $x < 0$ is defined by
\begin{align} \label{3.10}
\Psi_D(-x) &= \eta\, \Psi_D^*(x) \hspace{1cm} (\eta=\pm 1)
\end{align}
This implies that $\vphi_D(0)$ and $\chi_D(0)$ are both real $(\eta=1)$ or imaginary $(\eta=-1)$, which fixes two of the four real parameters in $c_1,c_2$. The remaining two parameters may be used to fix the norms of $\vphi_D(0)$ and $\chi_D(0)$, completely determining the solution for any $M_D$. When obtaining the Dirac equation as a limit of a bound state equation the value of $M_D$ must be supplied by the bound states.

\section{A numerical example} \label{secIV}

The bound state wave function \eq{3.4} can be reduced to the Dirac wave function \eq{3.9} for $m_2 \to \infty$, since the BSE \eq{2.16} reduces to the Dirac equation \eq{2.19}. One should be able to show this analytically by taking a limit of the $\kum(a,b,z)$ functions in \eq{3.4} for $a,b,z \propto m_2^2 \to \infty$. However, here I consider a numerical example.

\begin{figure} \centering
\includegraphics[width=1.\columnwidth]{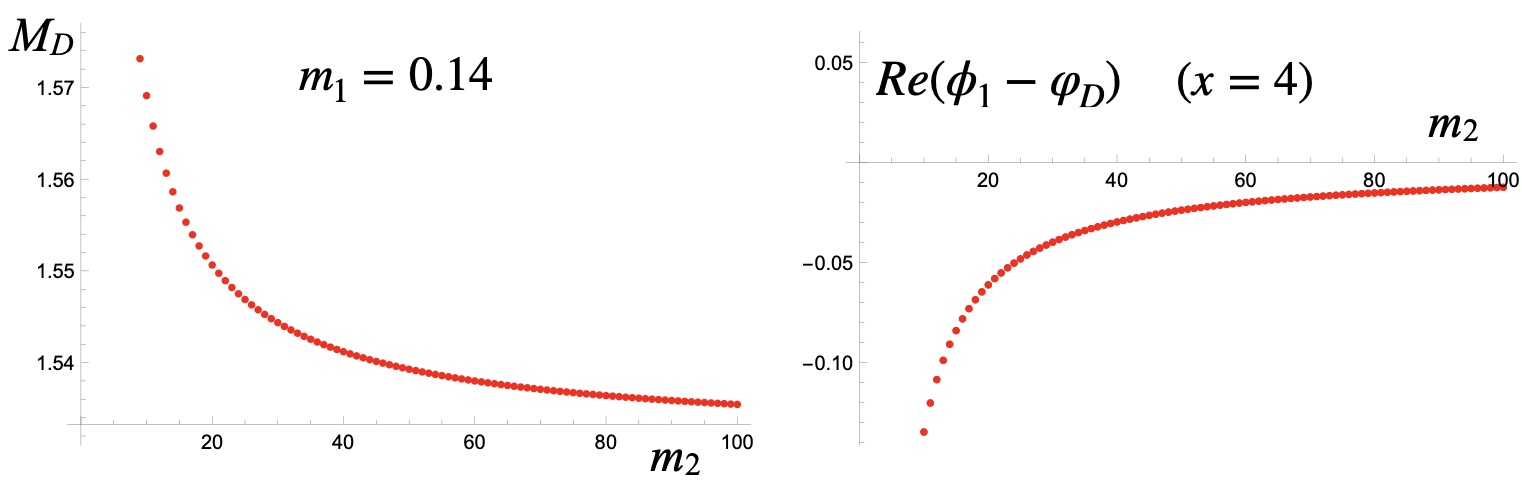}
\caption{Approach of the $P=0$ bound state wave function \eq{3.4} to the Dirac wave function \eq{3.9} with increasing $m_2$. All units are in terms of $V'=1$ and $m_1=0.14$. \ \textit{Left:} $M-m_2$ \eq{2.7} tending to $M_D$=1.532.\ \textit{Right:} $\mrm{Re}(\phi_1-\vphi_D)$ at $x=4.0$. \label{f1}}
\end{figure}

I take $m_1 = 0.14\sqrt{V'}$ as in \cite{Hoyer:2025cpf,Dietrich:2012un} and $P=0$. The parity in \eq{3.6} and \eq{3.10} is $\eta=1$ and the normalization is (arbitrarily) fixed by $\phi_1(0)=1$. Requiring $\mrm{Im}\,\phi_0(0)=0$ \eq{3.7} determines $M$ for each $m_2$. \fig{f1} (left) shows that $M-m_2$ approaches a fixed value $M_D$ as assumed in \eq{2.7}. Fitting a fourth order polynomial in $1/m_2$ to the numerical values of $M-m_2$ for $m_2 \leq 100\sqrt{V'}$ gives $M_D = 1.532\sqrt{V'}$\,. \fig{f1} (right) illustrates the convergence of the bound state and Dirac wave functions for increasing $m_2$ at $x=4.0/\sqrt{V'}$. The parameters $c_1,c_2$ of the Dirac solution \eq{3.9} were determined by $\vphi_D(0)=\phi_1(0)$ and $\chi_D(0)=-\phi_0(0)$ according to \eq{2.9} and \eq{3.8}, with $M_D = 1.532\sqrt{V'}$. The approach to the Dirac limit is consistent with being $\propto 1/m_2$ as in the analytic limit of the bound state equation \eq{2.6}.

\begin{figure} \centering
\includegraphics[width=.89\columnwidth]{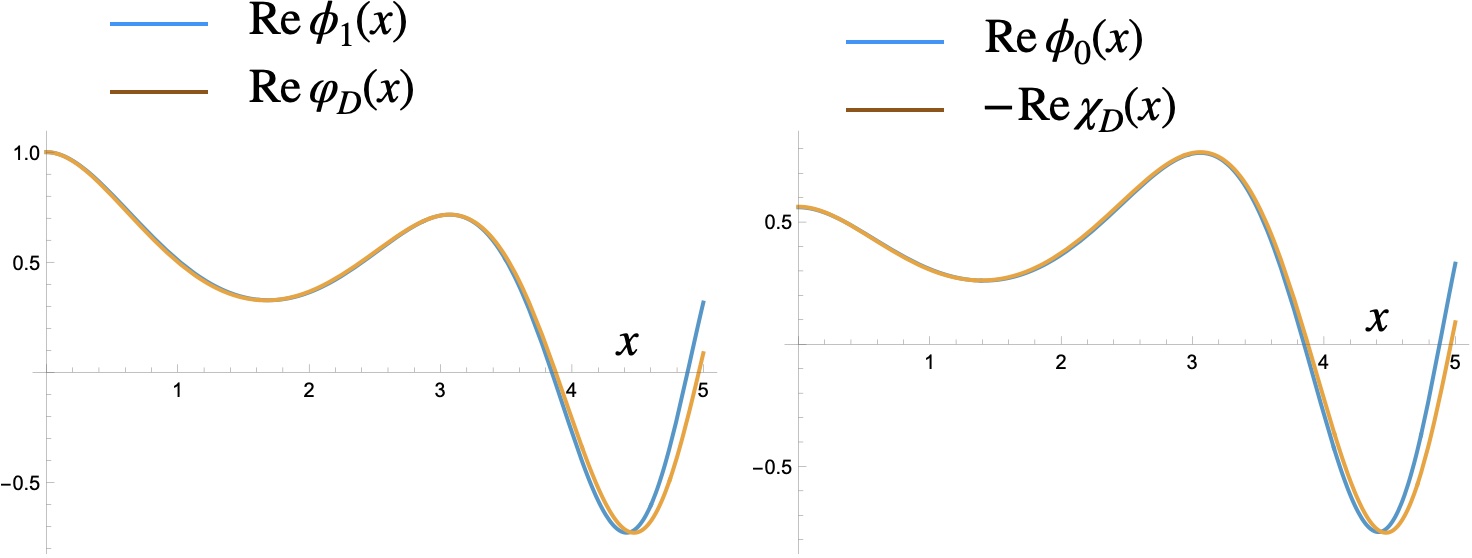}
\caption{Comparison of the bound state wave function \eq{2.6} ($m_1=0.14,\ m_2=20,\ M-m_2=1.551$) and the Dirac wave function \eq{3.9} $(m=0.14,\ M_D=1.532)$. All units are in terms of $V'=1$.  \ \textit{Left:} $\mrm{Re}\,\phi_1(x)$ (blue) and $\mrm{Re}\,\vphi_D(x)$ (brown).\ \textit{Right:} $\mrm{Re}\,\phi_0(x)$ (blue) and $-\mrm{Re}\,\chi_D(x)$ (brown). The imaginary parts agree at a similar level. \label{f2}}
\end{figure}

In \fig{f2} the real parts of the bound state and Dirac wave functions are compared as functions of $x$ at $m_2=20 \sqrt{V'}$. The imaginary parts agree similarly. The deviation at $x=5.0/\sqrt{V'}$ is commensurate with the ratio $V'x/m_2 = 0.25$.

\section{Discussion} \label{secV}

The wave function of a free, relativistic electron is given by the Dirac equation, which has exact Poincar\'e covariance. The covariance is maintained for electrons in an external field, when the electromagnetic potential $A^\mu(t,\xv)$ is transformed as a four-vector. However, a non-vanishing potential breaks translation invariance. An electron bound by a static Coulomb potential can be in an energy eigenstate, but not in an eigenstate of spatial momentum.

When the electron is a constituent of a bound state (with no external field) the coordinate frame is defined by the CM momentum of the bound state. If the companion constituent is heavy compared to the binding potential the electron dynamics in the rest frame should be given by the Dirac equation with a Coulomb potential. By boosting the bound state one may study the potential and the electron (Dirac) wave function in any frame.

Such a study is generally not practical for field theory bound states. 
However, QCD$_2$ at large $N_c$, first studied by 't Hooft \cite{tHooft:1974pnl}, is
sufficiently simple to allow an analytic study. The limit of a large number of colors suppresses $q\qb$ production, and there are no gluon constituents in $D=1+1$ dimensions. Hence the dynamics of the single $q\qb$ pair can be studied in a field theory context with exact Poincar\'e invariance. In canonical quantization and temporal ($A^0_a=0$) gauge the potential is instantaneous in time and linear in space, $V=V'|x|$.

In Section \ref{secII} I verified that the QCD$_2$ bound state equation for quarks with masses $m_1,m_2$ reduces to the Dirac equation for quark $m_1$ in the linear potential when $\sqrt{V'}/m_2 \to 0$. This holds in any frame of the bound state, with the $q\qb$ wave function Lorentz contracting as expected. Since the contraction is irrelevant for the single particle Dirac equation the light quark dynamics is, remarkably, frame independent.

In Section \ref{secIII} I presented the analytic solutions for the QCD$_2$ and Dirac wave functions, in terms of confluent hypergeometric functions $\kum(a,b,z)$. A striking feature of the Dirac equation with a linear potential is that the energy spectrum is continuous, like for plane waves \cite{Plesset:1930zz}. The negative energy components of the Dirac equation describe positrons, which are repelled by the potential and dominate at large $|x|$. The $q\qb$ wave function of QCD$_2$ may be singular at the value of $x$ where $V'|x|=E-P$. The singularity is absent only for discrete values of the bound state mass $M$. For $m_2 \to \infty$ the singular point moves to infinite $x$. Hence the Dirac wave function is regular, but only discrete values of the Dirac energy $M_D=M-m_2$ occur as the limit of $q\qb$ bound states.

In Section \ref{secIV} I studied the approach of the $q\qb$ bound state wave functions to the Dirac ones. Since the exact expressions are known it should be possible to take the $\sqrt{V'}/m_2 \to 0$ limit analytically in the $\kum(a,b,z)$ functions. I left this for future work and instead chose a numerical example in the rest frame. The difference $M-m_2$ converged to a constant value for the Dirac energy $M_D$, as shown in \fig{f1} (left). The difference between the $q\qb$ and Dirac wave functions also converged $\propto 1/m_2$ (\fig{f1}, right). The $x$-dependence of the wave functions were compared in \fig{f2} for $m_2 = 20\,\sqrt{V'}$, showing good agreement in the range where $V'|x| \ll m_2$.

In conclusion, the dynamics of a light quark bound to a heavy quark in
QCD$_2$ is described by the Dirac equation with a linear Coulomb potential in all frames, when using the method of \cite{Hoyer:2025cpf}.
This study can be extended to $D=3+1$ dimensions given a relativistic bound state framework. The approach described in Section VIII of \cite{Hoyer:2021adf} should provide this.

\begin{acknowledgments}
I thank Alexey Nefediev for discussions and for reading the manuscript. 
I am grateful for the support provided by the Physics Department of the University of Helsinki.
\end{acknowledgments}

\bibliography{refs.bib}

\begin{thebibliography}{13}%
\makeatletter
\providecommand \@ifxundefined [1]{%
 \@ifx{#1\undefined}
}%
\providecommand \@ifnum [1]{%
 \ifnum #1\expandafter \@firstoftwo
 \else \expandafter \@secondoftwo
 \fi
}%
\providecommand \@ifx [1]{%
 \ifx #1\expandafter \@firstoftwo
 \else \expandafter \@secondoftwo
 \fi
}%
\providecommand \natexlab [1]{#1}%
\providecommand \enquote  [1]{``#1''}%
\providecommand \bibnamefont  [1]{#1}%
\providecommand \bibfnamefont [1]{#1}%
\providecommand \citenamefont [1]{#1}%
\providecommand \href@noop [0]{\@secondoftwo}%
\providecommand \href [0]{\begingroup \@sanitize@url \@href}%
\providecommand \@href[1]{\@@startlink{#1}\@@href}%
\providecommand \@@href[1]{\endgroup#1\@@endlink}%
\providecommand \@sanitize@url [0]{\catcode `\\12\catcode `\$12\catcode
  `\&12\catcode `\#12\catcode `\^12\catcode `\_12\catcode `\%12\relax}%
\providecommand \@@startlink[1]{}%
\providecommand \@@endlink[0]{}%
\providecommand \url  [0]{\begingroup\@sanitize@url \@url }%
\providecommand \@url [1]{\endgroup\@href {#1}{\urlprefix }}%
\providecommand \urlprefix  [0]{URL }%
\providecommand \Eprint [0]{\href }%
\providecommand \doibase [0]{http://dx.doi.org/}%
\providecommand \selectlanguage [0]{\@gobble}%
\providecommand \bibinfo  [0]{\@secondoftwo}%
\providecommand \bibfield  [0]{\@secondoftwo}%
\providecommand \translation [1]{[#1]}%
\providecommand \BibitemOpen [0]{}%
\providecommand \bibitemStop [0]{}%
\providecommand \bibitemNoStop [0]{.\EOS\space}%
\providecommand \EOS [0]{\spacefactor3000\relax}%
\providecommand \BibitemShut  [1]{\csname bibitem#1\endcsname}%
\let\auto@bib@innerbib\@empty
\bibitem [{\citenamefont {Dirac}(1928{\natexlab{a}})}]{Dirac:1928hu}%
  \BibitemOpen
  \bibfield  {author} {\bibinfo {author} {\bibfnamefont {P.~A.~M.}\
  \bibnamefont {Dirac}},\ }\href {\doibase 10.1098/rspa.1928.0023} {\bibfield
  {journal} {\bibinfo  {journal} {Proc. Roy. Soc. Lond. A}\ }\textbf {\bibinfo
  {volume} {A117}},\ \bibinfo {pages} {610} (\bibinfo {year}
  {1928}{\natexlab{a}})}\BibitemShut {NoStop}%
\bibitem [{\citenamefont {Dirac}(1928{\natexlab{b}})}]{Dirac:1928ej}%
  \BibitemOpen
  \bibfield  {author} {\bibinfo {author} {\bibfnamefont {P.~A.~M.}\
  \bibnamefont {Dirac}},\ }\href {\doibase 10.1098/rspa.1928.0056} {\bibfield
  {journal} {\bibinfo  {journal} {Proc. Roy. Soc. Lond. A}\ }\textbf {\bibinfo
  {volume} {A118}},\ \bibinfo {pages} {351} (\bibinfo {year}
  {1928}{\natexlab{b}})}\BibitemShut {NoStop}%
\bibitem [{\citenamefont {Brodsky}(7494)}]{Brodsky:1971we}%
  \BibitemOpen
  \bibfield  {author} {\bibinfo {author} {\bibfnamefont {S.~J.}\ \bibnamefont
  {Brodsky}},\ }\href@noop {} {\bibfield  {journal} {\bibinfo  {journal}
  {Atomic physics and astrophysics. Vol.1. Brandeis University Summer Institute
  in Theoretical Physics}\ ,\ \bibinfo {pages} {91}} (\bibinfo {year} {1971,
  https://inspirehep.net/literature/67494})}\BibitemShut {NoStop}%
\bibitem [{\citenamefont {Gross}(1982)}]{Gross:1982nz}%
  \BibitemOpen
  \bibfield  {author} {\bibinfo {author} {\bibfnamefont {F.}~\bibnamefont
  {Gross}},\ }\href {\doibase 10.1103/PhysRevC.26.2203} {\bibfield  {journal}
  {\bibinfo  {journal} {Phys. Rev. C}\ }\textbf {\bibinfo {volume} {26}},\
  \bibinfo {pages} {2203} (\bibinfo {year} {1982})}\BibitemShut {NoStop}%
\bibitem [{\citenamefont {Neghabian}\ and\ \citenamefont
  {Gloeckle}(1983)}]{Neghabian:1983vm}%
  \BibitemOpen
  \bibfield  {author} {\bibinfo {author} {\bibfnamefont {A.}~\bibnamefont
  {Neghabian}}\ and\ \bibinfo {author} {\bibfnamefont {W.}~\bibnamefont
  {Gloeckle}},\ }\href {\doibase 10.1139/p83-014} {\bibfield  {journal}
  {\bibinfo  {journal} {Can. J. Phys.}\ }\textbf {\bibinfo {volume} {61}},\
  \bibinfo {pages} {85} (\bibinfo {year} {1983})}\BibitemShut {NoStop}%
\bibitem [{\citenamefont {J{\"a}rvinen}(2005)}]{Jarvinen:2004pi}%
  \BibitemOpen
  \bibfield  {author} {\bibinfo {author} {\bibfnamefont {M.}~\bibnamefont
  {J{\"a}rvinen}},\ }\href {\doibase 10.1103/PhysRevD.71.085006} {\bibfield
  {journal} {\bibinfo  {journal} {Phys. Rev.}\ }\textbf {\bibinfo {volume}
  {D71}},\ \bibinfo {pages} {085006} (\bibinfo {year} {2005})},\ \Eprint
  {http://arxiv.org/abs/hep-ph/0411208} {arXiv:hep-ph/0411208 [hep-ph]}
  \BibitemShut {NoStop}%
\bibitem [{\citenamefont {'t~Hooft}(1974)}]{tHooft:1974pnl}%
  \BibitemOpen
  \bibfield  {author} {\bibinfo {author} {\bibfnamefont {G.}~\bibnamefont
  {'t~Hooft}},\ }\href {\doibase 10.1016/0550-3213(74)90088-1} {\bibfield
  {journal} {\bibinfo  {journal} {Nucl. Phys. B}\ }\textbf {\bibinfo {volume}
  {75}},\ \bibinfo {pages} {461} (\bibinfo {year} {1974})}\BibitemShut
  {NoStop}%
\bibitem [{\citenamefont {Bars}\ and\ \citenamefont
  {Green}(1978)}]{Bars:1977ud}%
  \BibitemOpen
  \bibfield  {author} {\bibinfo {author} {\bibfnamefont {I.}~\bibnamefont
  {Bars}}\ and\ \bibinfo {author} {\bibfnamefont {M.~B.}\ \bibnamefont
  {Green}},\ }\href {\doibase 10.1103/PhysRevD.17.537} {\bibfield  {journal}
  {\bibinfo  {journal} {Phys. Rev. D}\ }\textbf {\bibinfo {volume} {17}},\
  \bibinfo {pages} {537} (\bibinfo {year} {1978})}\BibitemShut {NoStop}%
\bibitem [{\citenamefont {Hoyer}(2025)}]{Hoyer:2025cpf}%
  \BibitemOpen
  \bibfield  {author} {\bibinfo {author} {\bibfnamefont {P.}~\bibnamefont
  {Hoyer}},\ }\href {\doibase 10.1103/PhysRevD.111.074021} {\bibfield
  {journal} {\bibinfo  {journal} {Phys. Rev. D}\ }\textbf {\bibinfo {volume}
  {111}},\ \bibinfo {pages} {074021} (\bibinfo {year} {2025})},\ \Eprint
  {http://arxiv.org/abs/2501.18352} {arXiv:2501.18352 [hep-ph]} \BibitemShut
  {NoStop}%
\bibitem [{\citenamefont {Kalashnikova}\ and\ \citenamefont
  {Nefediev}(1999)}]{Kalashnikova:1997tb}%
  \BibitemOpen
  \bibfield  {author} {\bibinfo {author} {\bibfnamefont {Y.~S.}\ \bibnamefont
  {Kalashnikova}}\ and\ \bibinfo {author} {\bibfnamefont {A.~V.}\ \bibnamefont
  {Nefediev}},\ }\href@noop {} {\bibfield  {journal} {\bibinfo  {journal}
  {Phys. Atom. Nucl.}\ }\textbf {\bibinfo {volume} {62}},\ \bibinfo {pages}
  {323} (\bibinfo {year} {1999})},\ \Eprint
  {http://arxiv.org/abs/hep-ph/9711347} {arXiv:hep-ph/9711347} \BibitemShut
  {NoStop}%
\bibitem [{\citenamefont {Plesset}(1932)}]{Plesset:1930zz}%
  \BibitemOpen
  \bibfield  {author} {\bibinfo {author} {\bibfnamefont {M.~S.}\ \bibnamefont
  {Plesset}},\ }\href {\doibase 10.1103/PhysRev.41.278} {\bibfield  {journal}
  {\bibinfo  {journal} {Phys. Rev.}\ }\textbf {\bibinfo {volume} {41}},\
  \bibinfo {pages} {278} (\bibinfo {year} {1932})}\BibitemShut {NoStop}%
\bibitem [{\citenamefont {Hoyer}(2021)}]{Hoyer:2021adf}%
  \BibitemOpen
  \bibfield  {author} {\bibinfo {author} {\bibfnamefont {P.}~\bibnamefont
  {Hoyer}},\ }\href {\doibase 10.1007/978-3-030-79489-7} {\emph {\bibinfo
  {title} {{Journey to the Bound States}}}},\ SpringerBriefs in Physics\
  (\bibinfo  {publisher} {Springer},\ \bibinfo {year} {2021})\ \Eprint
  {http://arxiv.org/abs/2101.06721} {arXiv:2101.06721 [hep-ph]} \BibitemShut
  {NoStop}%
\bibitem [{\citenamefont {Dietrich}\ \emph {et~al.}(2013)\citenamefont
  {Dietrich}, \citenamefont {Hoyer},\ and\ \citenamefont
  {J{\"a}rvinen}}]{Dietrich:2012un}%
  \BibitemOpen
  \bibfield  {author} {\bibinfo {author} {\bibfnamefont {D.~D.}\ \bibnamefont
  {Dietrich}}, \bibinfo {author} {\bibfnamefont {P.}~\bibnamefont {Hoyer}}, \
  and\ \bibinfo {author} {\bibfnamefont {M.}~\bibnamefont {J{\"a}rvinen}},\
  }\href {\doibase 10.1103/PhysRevD.87.065021} {\bibfield  {journal} {\bibinfo
  {journal} {Phys. Rev.}\ }\textbf {\bibinfo {volume} {D87}},\ \bibinfo {pages}
  {065021} (\bibinfo {year} {2013})},\ \Eprint {http://arxiv.org/abs/1212.4747}
  {arXiv:1212.4747 [hep-ph]} \BibitemShut {NoStop}%
\end{thebibliography}%
\end{document}